\documentstyle[twoside,fleqn,espcrc2,fps]{article}

\title{How much are 2d Yukawa models similar to the Gross-Neveu models?%
\thanks{Talk presented by E. Focht at the ''Lattice '92'' conference,
        Amsterdam.}}

\author{A.K. De\address{Washington University, Department of Physics,
        St. Louis MO 63130, USA},
        E. Focht$^{\rm b,c}$,
        W. Franzki$^{\rm b,c}$
        and J. Jers\'ak\address{Institute for Theoretical Physics E,
        RWTH Aachen, Sommerfeldstr., 5100 Aachen,
        Germany}$^{\!\!\!,}$\address{HLRZ c/o KFA J\"ulich, P.O. Box 1913,
        5170 J\"ulich,
        Germany}
       }

\begin{document}

\begin{abstract}
We present numerical evidence that the 2d Yukawa models with strong
quartic selfcoupling of the scalar field have the same phase structure
and are asymptotically free in the Yukawa coupling like the
Gross-Neveu models.
\end{abstract}

\maketitle


\newcommand{\vers}{\today}
%

%

\newcommand{\bibi}{\bibitem}
\newcommand{\un}{1\!\!1}
\newcommand{\half}{\frac{1}{2}}
\newcommand{\NN}{N\!\!|}
\newcommand{\al}{\alpha}
\newcommand{\bt}{\beta}
\newcommand{\lag}{\langle}
\newcommand{\rag}{\rangle}
\newcommand{\gm}{\gamma}
\newcommand{\pl}{\partial}
\newcommand{\Gm}{\Gamma}
\newcommand{\Dm}{D_{\mu}^+}
\newcommand{\Dbm}{D_{\mu}^-}
\newcommand{\Pm}{\partial_{\mu}^+}
\newcommand{\Pbm}{\partial_{\mu}^-}
\newcommand{\dl}{\delta}
\newcommand{\ep}{\epsilon}
\newcommand{\vv}{\langle \Phi \rangle}
\newcommand{\vvs}{\langle \Phi_{st} \rangle}
\newcommand{\pup}{\langle \Phi^{\dagger} U \Phi \rangle}
\newcommand{\vep}{\varepsilon}
\newcommand{\vst}{v_{st}}
\newcommand{\hst}{h_{st}}
\newcommand{\zt}{\zeta}
\newcommand{\et}{\eta}
\newcommand{\th}{\theta}
\newcommand{\kp}{\kappa}
\newcommand{\lm}{\lambda}
\newcommand{\lmp}{{\lambda^{\prime}}}
\newcommand{\lmb}{\overline{\lambda}}
\newcommand{\lmbp}{{\overline{\lambda}^{\prime}}}
\newcommand{\rh}{\rho}
\newcommand{\sg}{\sigma}
\newcommand{\ta}{\tau}
\newcommand{\ph}{\phi}
\newcommand{\vr}{\varphi}
\newcommand{\ch}{\chi}
\newcommand{\ps}{\psi}
\newcommand{\om}{\omega}
\newcommand{\phat}{{\hat{p}}^2}
\newcommand{\umxl}{U_{x \mu}^L}
\newcommand{\umxr}{U_{x \mu}^R}
\newcommand{\phd}{\phi^{\dagger}}
\newcommand{\Ps}{\Psi}
\newcommand{\Ph}{\Phi}
\newcommand{\Phd}{\Phi^{\dagger}}
\newcommand{\Om}{\Omega}
\newcommand{\Psb}{\overline{\Ps}}
\newcommand{\psb}{\overline{\ps}}
\newcommand{\ybb}{\overline{y}}
\newcommand{\Zb}{\overline{Z}}
\newcommand{\cM}{{\cal M}}
\newcommand{\Dsl}{D\!\!\!\!/}
\newcommand{\Dmu}{D_{\mu}}
\newcommand{\dmu}{\partial_{\mu}}
\newcommand{\gmmu}{\gamma_{\mu}}
\newcommand{\dsl}{\partial \!\!\!/}
\newcommand{\PbP}{\langle\Psb \Ps\rangle}
\newcommand{\PPP}{\lag \Psb\Ph\Ps \rag}
\newcommand{\mn}{m_F^{(n)}}
\newcommand{\mc}{m_F^{(c)}}
\newcommand{\mdn}{m_D^{(n)}}
\newcommand{\mdc}{m_D^{(c)}}
\newcommand{\En}{E_F^{(n)}}
\newcommand{\Ec}{E_F^{(c)}}
\newcommand{\Edn}{E_D^{(n)}}
\newcommand{\Edc}{E_D^{(c)}}
\newcommand{\mpr}{m^{\prime}}
\newcommand{\mpp}{m^{\prime \prime}}
\newcommand{\mdp}{m_D^{\prime}}
\newcommand{\Pp}{\Psi^{\prime}}
\newcommand{\Pc}{\Psi^{(c)}}
\newcommand{\Pn}{\Psi^{(n)}}
\newcommand{\psn}{\psi^{(n)}}
\newcommand{\psch}{\psi^{(c)}}
\newcommand{\psbn}{\overline{\psi}^{(n)}}
\newcommand{\psbch}{\overline{\psi}^{(c)}}
\newcommand{\Ppp}{\Psi^{\prime \prime}}
\newcommand{\Ppb}{\overline{\Psi}^{\prime}}
\newcommand{\Pnb}{\overline{\Psi}^{(n)}}
\newcommand{\Pcb}{\overline{\Psi}^{(c)}}
\newcommand{\Pppb}{\overline{\Psi}^{\prime \prime}}
\newcommand{\mdpp}{m_D^{\prime \prime}}
\newcommand{\Mo}{M_{off}^{\prime}}
\newcommand{\INP}{[{\cal M}(1)+\dsl P_R]^{-1}}
\newcommand{\SP}{S_0}
\newcommand{\SPP}{S_0^{\prime \prime}}
\newcommand{\INQ}{\dsl ^{-1} {\cal M} (\Phi )}
\newcommand{\mb}{\overline{m}}
\newcommand{\yb}{\overline{y}}
\newcommand{\chb}{\overline{\chi}}
\newcommand{\hmu}{\hat{\mu}}
\newcommand{\ra}{\rightarrow}
\newcommand{\Ra}{\Rightarrow}
\newcommand{\be}{\begin{equation}}
\newcommand{\ee}{\end{equation}}
\newcommand{\bea}{\begin{eqnarray}}
\newcommand{\eea}{\end{eqnarray}}

\newcommand{\eq}{\ref}
\newcommand{\beq}{\begin{equation}}
\newcommand{\eeq}{\end{equation}}
\newcommand{\cc}{\cite}
\newcommand{\lb}{\label}
\newcommand{\gsim}{\stackrel{>}{\sim}}  
\newcommand{\lsim}{\stackrel{<}{\sim}} 
\newcommand{\ras}{\stackrel{s \rightarrow +\infty}{\rightarrow}}
\newcommand{\SU}{G$_L \otimes $G$_R\;$}
\newcommand{\SL}{(2)$_L$}
\newcommand{\SR}{SU(2)$_R$}
\newcommand{\SKU}{SU(2)$_L \otimes $U(1)$_{Y}$}
\newcommand{\UY}{U(1)$_{Y}$}
\newcommand{\Ww}{Wechselwirkung }
\newcommand{\os}{O(4)-sym\-me\-tri\-sche $\Phi^{4}$-The\-o\-rie }
\newcommand{\oz}{Z(2)-sym\-me\-tri\-sche $\Phi^{4}$-The\-o\-rie }
\newcommand{\osn}{O(4)-sym\-me\-tri\-schen $\Phi^{4}$-The\-o\-rie }
\newcommand{\ozn}{Z(2)-sym\-me\-tri\-schen $\Phi^{4}$-The\-o\-rie }
\newcommand{\op}{$\Phi^{4}$-The\-o\-rie }
\newcommand{\OH}{$O_{1}(\vec{x},t)=Re \sum_{\mu}\Phi_{x}^{*}U_{x,\mu}
\Phi_{x+\mu}$}
\newcommand{\OV}{$O_{2}(\vec{x},t)=Im \sum_{\mu}\Phi_{x}^{*}U_{x,\mu}
\Phi_{x+\mu}$}
\newcommand{\OP}{$O_{3}(\vec{x},t)=Im\sum_{\mu ,\nu} U_{P}$}
\newcommand{\OB}{$O_{4}(\vec{x},t)=Re\sum_{\mu ,\nu} U_{P}$}
\newcommand{\OHS}{$O_{1}(\vec{x},t)= \sum_{\mu}Sp(\Phi_{x}^
{\dagger}U_{x,\mu}\Phi_{x+\mu})$}
\newcommand{\OVS}{$O_{2}(\vec{x},t)= \sum_{\mu}Sp(\Phi_{x}^
{\dagger}U_{x,\mu}\Phi_{x+\mu}\sigma)$}
\newcommand{\OS}{$O_{1}(\vec{x},t)=\Phi_{s,x}$}
\newcommand{\OK}{$O_{2}(\vec{x},t)=Re \sum_{\mu} Sp \Phi_{x}^{\dagger}
                                                    \Phi_{x+\mu}$}
\newcommand{\OG}{$O_{3}(\vec{x},t)=\Phi^{\alpha}_{x}$}
\newcommand{\OGP}{$O_{4}(\vec{x},t)=\Phi_{\perp,x}$}
\newcommand{\OJ}{$O_{2}(\vec{x},t)=\Phi_{j,x}$}
\newcommand{\OGJ}{$O_{3}(\vec{x},t)=\Phi_{\perp,x}$}
\newcommand{\k}{$\kappa$ }
\newcommand{\w}{Wechselwirkungsfreiheit }
\newcommand{\Va}{Valenzn\"aherung }
\def \3{\ss}
\newcommand{\T}{Trivialit\"at }
\newcommand{\ov}{\overline}
\newcommand{\bo}{[}
\newcommand{\bc}{]}
\newcommand{\PLL}{\lag \Pn_L \Pnb_L \rag}
\newcommand{\PRR}{\lag \Pn_R \Pnb_R \rag}
\newcommand{\PLR}{\lag \Pn_L \Pnb_R \rag}
\newcommand{\PRL}{\lag \Pn_R \Pnb_L \rag}
\newcommand{\PPLL}{\lag \Ppp_L \Pppb_L \rag}
\newcommand{\PPRR}{\lag \Ppp_R \Pppb_R \rag}
\newcommand{\PPLR}{\lag \Ppp_L \Pppb_R \rag}
\newcommand{\PPRL}{\lag \Ppp_R \Pppb_L \rag}
\newcommand{\CGT}{\chi GT}

\newcommand{\GN}{Gross--Neveu{ }}
\newcommand{\MC}{Monte Carlo{ }}
\newcommand{\tr}{\mathop{\rm Sp} \nolimits}
\newcommand{\pht}{{\widetilde{\ph}}}
\newcommand{\vrd}{\varphi^{\dagger}}
\newcommand{\vrs}{\varphi^{*}}
\newcommand{\yt}{\widetilde{y}}
\newcommand{\Real}{I\!\!R}
\newcommand{\Lm}{\Lambda}
\newcommand{\psla}{/ \!\!\!p}
\newcommand{\ksl}{k \!\!\!/}
\newcommand{\aleq}{\mbox{}^{\small <}_{\small \sim}}
\newcommand{\ageq}{\mbox{}^{\small >}_{\small \sim}}

\newcommand{\Yu}{Y$_2$\ }
%
%
%

\section{Yukawa models in 2 dimensions}

The 2d Yukawa models (\Yu) with chiral Z(2) or U(1) symmetries
are natural extensions of the usual or chiral 2d Gross-Neveu
(GN) models, respectively.
Starting from the auxiliary field representation of the 4-fermion coupling,
one can add both the kinetic term and a self-interaction
of this field $\phi$ into the GN action.
On the lattice the Z(2) symmetric \Yu action is then
(we introduce $N_F/4$ ``naive'' Dirac fermion fields $\psi^\al$)
\bea
S & = & -2\kp\sum_{x,\mu}\ph_x\ph_{x+\mu}+\sum_x\ph_x^2+
          \lm\sum_x(\ph_x^2-1)^2 \nonumber \\
& + & \sum_{x,\al} \psb_x^\al\dsl\ps_x^\al
      + y\sum_{x,\al}\psb_x^\al\ph_x\ps_x^\al~.
\lb{ACTION}
\eea
The $\phi^4$ scalar selfcoupling has been chosen, out of many
possibilities in 2d, for the sake of analogy with the 4d theories.
For $\kp = \lm = 0$ the Z(2) GN model is obtained if
the Yukawa coupling $y$ is related to the usual GN coupling $g$ by
$y = \sqrt{2}g$.

By choosing the above hopping parameter $\kp$ formulation of the scalar
field sector
the kinetic term can be turned on or off {\em gradually},
elucidating the smoothness of the transition from the auxiliary
to a dynamical field.
The relationship
\beq
y_0 = \frac{y}{a\sqrt{2\kp}}
\lb{Y0}
\eeq
between the Yukawa coupling $y_0$ used in continuum and $y$
is singular at $\kp =0$, however.
Another virtue of the hopping parameter formulation is that
spin models with the Z(2) or U(1) symmetry
(the Ising or the XY models, respectively) are easily recovered
at $y=0$ and $\kp$ finite
by choosing $\lm = \infty$.
The \Yu models thus interpolate between the GN and spin models.

\section{Expected scaling properties}

Motivated by the recent discussion of a relationship between
the Nambu--Jona-Lasinio type four-fermion theories and the Standard
Model \cc{HaHa91,Zi91,BaHi90}
we address here the question in which regions of the parameter space
the \Yu models still possess the most cherished properties of the GN
models, namely the asymptotic freedom at $y \ra 0$, the fermion mass
generation and, in the case of the Z(2) model, the symmetry breaking
\cc{GrNe74,Wi78b}.
In the GN models these properties are derived by means of the 1/$N_F$
expansion.

For \Yu models this expansion is applicable
for $\lm$ = O(1/$N_F$) \cc{HaHa91,Zi91}
and here the same results as for the GN models are found.
In particular, as long as $\kp < \kp_c(\lm = 0) = 1/4$,
the fermion mass $am_F$ is expected to scale as
\beq
am_F \propto \exp\left[-\frac{1}{2\bt_0}
\frac{a^2m^2}{Z}\frac{1}{y^2} \right].
\lb{ASY}
\eeq
For $\lm=0$ we have
\beq
      m^2 = m_0^2 = \left(1-\frac{\kp}{\kp_c(0)}\right) \frac{1}{a^2\kp}
      ~,~~ Z = \frac{1}{2\kp}~,
\lb{M}
\eeq
with $m_0$ being the scalar field mass at $\lm = y = 0$.

For large $\lm$ the 1/$N_F$ expansion is \`a priori not applicable,
not to speak of the perturbation theory.
Nevertheless, M.A.~Stephanov suggested \cc{St92} that in this case
the mean field (MF) theory could be a good guide.
An observation of long range ferromagnetic couplings
in the effective scalar field theory at $y > 0$ by E.~Seiler \cc{Se92}
supports the applicability of the MF theory.
The resulting expectations are \cc{DeFo93b}:

{\it (i)}~~~ The \Yu models at $\lm = \infty$
or large possess at arbitrarily small $y$ only the broken symmetry
(in the Z(2) case) or the spin-wave (in the U(1) case) phase,
whereas the symmetric or vortex phase present in the $y=0$ case
in the pure scalar theory is absent at $y > 0$.

{\it (ii)}~~~ The fermion mass and the magnetization
$y \langle \phi \rangle$ scale according to eq.~(\eq{ASY}).
The coefficient $m^2$ is for $\kp>0$ the squared scalar
mass at $y=0$ and $Z$ its wave function renormalization constant.
In general, $Z/m^2$ could be replaced by the scalar field propagator
at zero momentum, i.e. the susceptibility.
This should hold for any $\kp < \kp_c(\lm)$ where $\kp_c(\lm)$
is the line of critical points in the pure scalar theory at $y=0$.

Thus according to the MF theory the asymptotic freedom and the other
mentioned properties of the GN models might occur also for large
$\lm$ in the \Yu models.

\section{Methods of data analysis}

We have tried three methods:\\
{\it (i)}~~~ The {\bf asymptotic scaling law}, which could
be used to fit the data, is as (\eq{ASY}) multiplied by
$y^{-\bt_1/\bt_0^2}$.
Here $\bt_0$ and $\bt_1$ are the $\bt$-function coefficients
which in the GN limit have the perturbative values \cc{We85,De88}
\beq
\bt_0=\frac{N_F-1}{2\pi} \quad,\qquad
\bt_1=-\frac{N_F-1}{(2\pi)^2}
\lb{BT_Z2}
\eeq
in the model with Z(2) symmetry and
\beq
\bt_0=\frac{N_F}{2\pi} \quad,\qquad
\bt_1=-\frac{N_F}{2\pi^2}
\lb{BT_U1}
\eeq
in the U(1) case.
This method does not take the finite volume
effects into account.
Nevertheless, one can roughly determine the values
of the coefficient $\bt_0$ and compare with the values
(\eq{BT_Z2}) and (\eq{BT_U1}).
The 2-loop contribution given by $\bt_1$ is less important than
the finite size effects.

{\it (ii)}~~~ A more appropriate way of analyzing
the data is provided by a
{\bf modified gap equation on finite lattices},
\beq
\frac{a^2m^2}{Z}\frac{1}{y^2}
=\frac{\pi\bt_0}{V}
\sum\limits_{ \{p\} }
\frac{1}{\sum\limits_\mu\sin^2p_\mu+(am_F/s )^2}
\lb{GE}
\eeq
where the sum is performed over the fermion momenta
with one periodic and one antiperiodic boundary condition
on the $L^2$ lattices.
Thanks to the IR divergence it leads for small $am_F$
in the infinite volume limit to
the scaling law (\eq{ASY}).

The coefficient $\bt_0$ is considered as a free parameter in order to
allow for its
possible deviation
from the
perturbative value.
The parameter $s$ takes into account the fact that the
gap equation determines the value
of the mass gap with insufficient precision. In order to fit the data
it is necessary to treat the mass gap as a free parameter.

It would be interesting to determine the mass gap and compare
it with the recently obtained exact results for the GN models
\cc{FoNi91a,FoNa92}.
With our present accuracy it could be only estimated to be
roughly consistent with these results.

We note that one could take the finite size effects into account
also beyond the leading $1/N_F$ order \cc{BeLa90,CaCu89}.
As we do not expect this expansion to be applicable
for large $\lm$, we do not attempt such refinements.

Once the parameters in the gap equation have been determined
by a fit, the quantity $h(\kp,\lm)$,
\beq
   h(\kp,\lm)=\frac{1}{2\bt_0}\frac{a^2m^2}{Z}\Ra
   am_F \propto \exp\left[-\frac{h(\kp,\lm)}{y^2}\right]
\lb{H}
\eeq
has been extracted.

{\it (iii)}~~~ We have tried to fit the data also by an alternative
to the essential singularity at $y=0$, namely by a hypothetical
power law behavior
\beq
              m_F=a(y-y_c)^b.
\lb{POWER}
\eeq

\section{Results at $\lm = 0$}

To gain experience we have first performed simulations at $\lm=0$
in the interval $-0.1 \le \kp < \kp_c(\lm = 0) = 1/4$.
We have determined in both models the $y$-dependence of $am_F$ and of
$y\lag\phi\rag$ in the Z(2) model
at fixed
values of $\kp$ on lattices of size 16$^2$ - 64$^2$.
The following observations are useful for the study of
the models at large $\lm$:

{\it (i)}~~~ The data analysis by means of the asymptotic law
(\eq{ASY})
both without and with the 2--loop correction
is possible if those points at small $y$, which show finite size effects,
are excluded (fig.~1). However, the obtained values of $\bt_0$ change with
lattice size, when lower values of $am_F$ can be taken into account
on larger lattices,
so that one probably does not see the true asymptotic behaviour.

{\it (ii)}~~ The gap equation (\eq{GE}) can describe the data obtained
on different lattices consistently by means of one set of parameter values,
including $\bt_0$ (fig.~1).
The onset of finite size effects at small $am_F$, as well as the
general tendency of the data at large $y$ are well reproduced.
This analysis is superior to that by means of (\eq{ASY}).

{\it (iii)}~~ The fermion masses in both models and
$y\langle \phi \rangle$
in the Z(2) model behave in a very analogous way.
The magnetization in the U(1) case is present on finite lattices, but
it shows a significant size dependence consistent with its vanishing
in the infinite volume.
Thus we observe the dynamical fermion mass generation taking place
in spite of the absence of spontaneous symmetry breaking in 2d
\cc{Wi78b}.

{\it (iv)}~~ The $\kp$-dependence is consistent with the expected one,
eq.~(\eq{M}). This is demonstrated in fig.~2, where
$h(\kp,0)$ is shown.
The values of $\bt_0$ obtained at different $\kp$ are for both
models consistent with the values (\eq{BT_Z2}) and (\eq{BT_U1})
within 10 \%.

{\it (v)}~~ The power law formula (\eq{POWER})
can fit the data well for any
given lattice size. The parameters $b$ and $y_c$ depend strongly
on the lattice size, however.
This makes an algebraic singularity less probable than the essential
one, but a reliable exclusion of the former on the basis of data alone
seems very difficult.
So one should be cautious at large $\lm$, when the analytic information
on the type of singularity is less reliable than at $\lm = 0$.

\section{Results at large $\lm$}

Going to large $\lm$ we have simulated both models at $\lm = 0.5$
and the U(1) model at $\lm = \infty$ in the interval
$-0.3 \le \kp < \kp_c(\lm)$.
The most important results are:

{\it (i)}~~~ The $y$-dependence of the fermion mass,
including the onset of the finite size effects,
is described by the gap equation (\eq{GE})
nearly as well as in the $\lm = 0$ cases.
This is demonstrated in fig.~3 for the U(1) model at $\lm = \infty$
and $\kp = 0$ on 16$^2$ - 64$^2$ lattices.

{\it (ii)}~~~ The power law behaviour (\eq{POWER}) is not excluded but,
similarly to $\lm=0$, disfavored by the strong dependence of $y_c$
and $b$ on the lattice size. However, the finite size scaling analysis
based on the Ansatz (\eq{POWER}) has yet to be done.

{\it (iii)}~~~ These facts lead us to the tentative conclusion
that at large $\lm$, including $\lm = \infty$, only the broken
symmetry (Z(2) model) or spin wave (U(1) model) phase is present
at arbitrary small $y$.

{\it (iv)}~~~ Except at $\kp=0$, we do not yet have an independent
determination of $m^2/Z$ for large $\lm$. Therefore we cannot yet
determine the values of $\bt_0$ and give the results
in form of the values of the coefficient $h(\kp,\lm)$. As an
example we show $h(\kp,\infty)$ for the U(1) model in fig.~4
(here $\kp_c(\infty)\approx0.56$). We hope
to determine $\bt_0$ at large $\lm$ for various $\kp$ in the near
future. At present our estimates indicate that its values are roughly
consistent with $\bt_0$ in the GN cases, eqs. (\eq{BT_Z2}) and (\eq{BT_U1}).

Thus we have found some evidence that the \Yu models
with Z(2) and U(1) chiral symmetries
with a strong $\phi^4$ interaction behave for $\kp < \kp_c(\lm)$
as the GN models of the same symmetry,
in particular they are asymptotically free. The transition from
the GN to the spin model universality classes takes place probably at
small $y$ and $\kp\approx\kp_c(\lm)$.

\section{Acknowledgements}

We thank E.~Seiler and M.A.~Stephanov for helpful suggestions and
A.~Hasenfratz, P.~Hasenfratz, R.~Lacaze, F.~Niedermayer and M.M.~Tsypin
for valuable discussions.
The calculations have been performed on the CRAY Y-MP of HLRZ J\"ulich.
This work has been supported by Deutsches Bundesministerium
f\"ur Forschung und Technologie and by Deutsche Forschungsgemeinschaft.

\vspace{-2cm}
%
%
%
\begin{figure}[b]
\centerline{
 \vspace{-1cm}
 \epsfxsize=5.2cm
 \epsfbox{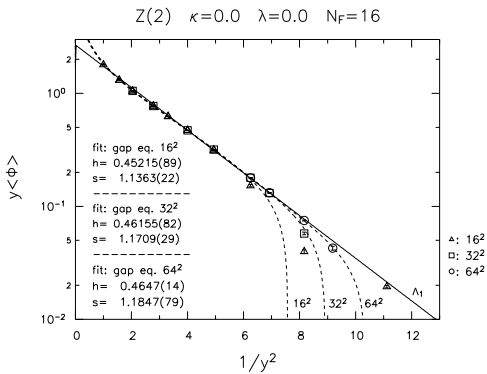}
}
\caption{ \noindent {\em  The $1/y^2$ dependence of $y\langle \phi \rangle$
in the Z(2) GN model ($\kp=\lm=0$) with fits by eq.~(\protect\eq{GE}).
The straight line is a fit by eq.~(\protect\eq{ASY}).}}
\label{GN_exp_gap}
\end{figure}
%
%
%
%
\begin{figure}
\centerline{
\vspace{-1cm}
\epsfxsize=5.2cm
\epsfbox{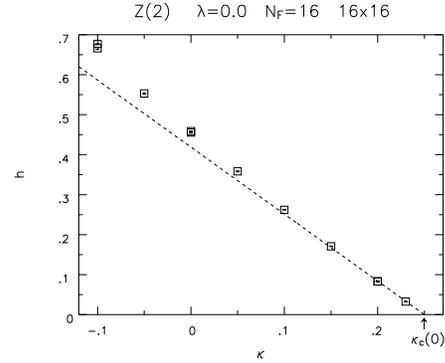}
}
\caption{ \noindent {\em  The $\kp$ dependence of the coefficient $h(\kp,0)$
in the Z(2) model at $\lm=0$ is consistent with eq.~(\protect\eq{M})
(dashed line).}}
\label{H_KP_L0_Z2}
\end{figure}
%
%
%
%
%
\begin{figure}
\centerline{
\vspace{-1cm}
\epsfxsize=5.2cm
\epsfbox{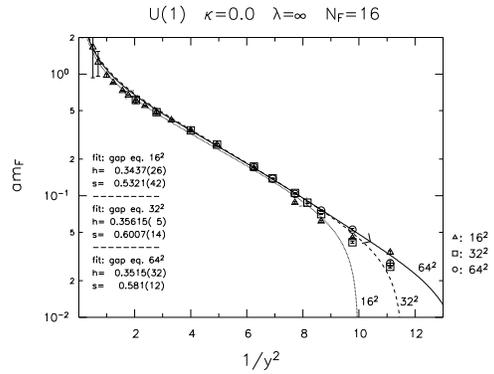}
}
\caption{ \noindent {\em  The $1/y^2$ dependence of $am_F$
in the U(1) \Yu model with $N_F= 16$ at $\lm = \infty$
with fits by eq.~(\protect\eq{GE}).}}
\label{ui_exp_1}
\end{figure}
%
%
%
%
%
\begin{figure}
\centerline{
\vspace{-1cm}
\epsfxsize=5.2cm
\epsfbox{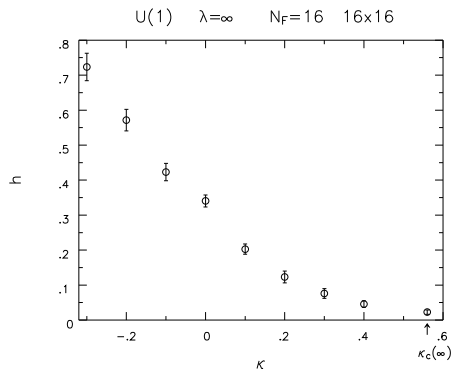}
}
\caption{ \noindent {\em  The $\kp$ dependence of the coefficient
$h(\kp,\infty)$
in the U(1) model at $\lm=\infty$.}}
\label{H_KP_LINF_U1}
\end{figure}

\end{document}